\documentclass[preprint,aps,prd,nofootinbib]{revtex4-1}

    \usepackage{graphicx}
    \usepackage{subfig, tabularx, afterpage,xifthen}
    \usepackage{amsmath,amsfonts,amssymb,mathtools}
    \usepackage{overpic}
    \pdfoutput=1 
    
    \newcommand{\fr}[2]{\frac{#1}{#2}}

    \newcommand{\beq}{\begin{equation}}
    \newcommand{\eeq}{\end{equation}}
    \newcommand{\ba}{\begin{eqnarray}}
    \newcommand{\ea}{\end{eqnarray}}
    \newcommand{\bry}{\begin{array}}
    \newcommand{\ery}{\end{array}}
    \newcommand{\bit}{\begin{itemize}}
    \newcommand{\eit}{\end{itemize}}
    \newcommand{\ben}{\begin{enumerate}}
    \newcommand{\een}{\end{enumerate}}
    \newcommand{\btab}{\begin{tabular}}
    \newcommand{\etab}{\end{tabular}}
    \newcommand{\bfig}{\begin{figure*}[tb]\centering}
    \newcommand{\efig}{\end{figure*}}
    \def\bwtxt{\begin{widetext}}
    \def\ewtxt{\end{widetext}}
    \def\bop{\begin{overpic}}
    \def\eop{\end{overpic}}
    \def\bpm{\begin{pmatrix}}
    \def\epm{\end{pmatrix}}
    \def\bbm{\begin{bmatrix}}
    \def\ebm{\end{bmatrix}}

    \def\alsb{{\bar{\al_s}}}
    \def\lag{{\cal L}}

    \def\lqcd{\Lambda_{\rm QCD}}
    \def\nnlo{N$^2$LO}

    \def\d{{\rm d}}
    \newcommand{\lo}[1][]{\ifthenelse{\isempty{#1}}{LO}{LO$_{#1}$}}
    \newcommand{\nlo}[1][]{\ifthenelse{\isempty{#1}}{NLO}{NLO$_{#1}$}}

    \def\msbar{\overline{\rm MS}}
    \def\zb{o\!\!\!{\scriptscriptstyle/}}
    \def\cp{$\mathit{CP}$}

        \newcommand{\al}{\alpha}
        
        \newcommand{\ga}{\gamma}
        \newcommand{\de}{\delta}
        \newcommand{\ep}{\epsilon}

        \newcommand{\ka}{\kappa}
        \newcommand{\la}{\lambda}
        
        \newcommand{\si}{\sigma}

        \newcommand{\Ga}{\Gamma}
        \newcommand{\De}{\Delta}
        \newcommand{\La}{\Lambda}

        \newcommand{\ptl}{\partial}

    \def\nn{\nonumber \\}
    
    \newcommand{\non}{\nonumber}

    \def\OMIT#1{{}}

    \makeatletter
    \def\fmslash{\@ifnextchar[{\fmsl@sh}{\fmsl@sh[0mu]}}
    \def\fmsl@sh[#1]#2{%
       \mathchoice
         {\@fmsl@sh\displaystyle{#1}{#2}}%
         {\@fmsl@sh\textstyle{#1}{#2}}%
         {\@fmsl@sh\scriptstyle{#1}{#2}}%
         {\@fmsl@sh\scriptscriptstyle{#1}{#2}}}
    \def\@fmsl@sh#1#2#3{\m@th\ooalign{$\hfil#1\mkern#2/\hfil$\crcr$#1#3$}}
    \makeatother

    \newcommand{\ord}{{\mathcal O}}
    
    \newcommand{\oa}{\ord(\alpha_s)}
    \newcommand{\oat}{\ord(\alpha_s^2)}
    \def\ordt{$\ord(\tau)$}
    
    \renewcommand{\log}{\ln} 

    \def\nslash{\fmslash{n}}
    \def\nbslash{\bar n\!\!\!\slash\,\,\,}
    \def\bnslash{{\fmslash{\nb}}}

    \def\nb{{\bar{n}}}
    
    \def\ncol{$n$-collinear}
    \def\nbcol{$\bar n$-collinear}
    
    \def\qcd{\text{\tiny QCD}}
    
    \def\scet{\text{\tiny SCET}}
    \def\ca{C_A}
    \def\cf{C_F}
    \def\cah{\fr{\ca}{2}}

    \def\dereg{$\de$-regulator}
    \def\PlusF#1{\left[#1\right]_+}

    \def\nbk{{k^-}}
    \def\ub{{\bar u}}
    \def\vb{{\bar v}}
    \def\lq{\log\fr{-Q^2}{\mu^2}}
    \def\alsb{\left(\fr{\al_s}{4\pi}\right)}

    \def\t{{\hat t}}
    \def\xnb{{x_\nb}}
    \def\xn{{x_n}}
    \def\xnty{{x_n^\infty}}
    \def\xnbty{{x_\nb^\infty}}
    \def\xsnty{{x_{us}^{\infty_n}}}
    \def\xsnbty{{x_{us}^{\infty_\nb}}}
    \def\wn#1{{W_n^{(\bf #1)}}{}}
    \def\wnb#1{{W_\nb^{(\bf #1)}}{}}
    \def\yn#1{{Y_n^{(\bf #1)}}{}}
    \def\ynb#1{{Y_\nb^{(\bf #1)}}{}}

    \def\op#1{O_2^{(#1)}{}}
    \def\opft#1{\tilde{O}_2^{(#1)}{}}
    \def\wc#1{{C_2^{(#1)}}}
    \def\wcft#1{{\tilde{C}_2^{(#1)}}}
    \def\ct#1{Z_{2(#1)}}
    \def\ad#1{{\ga_{2(#1)}}}
    \def\Gai#1{\Ga_{(#1)}}
    
    \def\gai#1{\ga_{(#1)}}

    \def\intfm{\kappa_\epsilon \int {\rm d}^d }
    \newcommand{\dfthat}[1][]{\ifthenelse{\isempty{#1}}{\fr{\d\t}{(2\pi)}\,}{\fr{\d\t_#1}{(2\pi)}\,}}   
    \newcommand{\dft}[1][]{\ifthenelse{\isempty{#1}}{\fr{\d t}{(2\pi)}\,}{\fr{\d t_#1}{(2\pi)}\,}}

    \def\labelscet{Bauer:2000ew,Bauer:2000yr,Bauer:2001ct,Bauer:2001yt,Bauer:2002nz} 

\begin{document}

\title{Renormalization of Subleading Dijet Operators in Soft-Collinear Effective Theory}
\author{Simon M. Freedman}
\email{sfreedma@physics.utoronto.ca}
\affiliation{Department of Physics, University of Toronto, 60 St.\ George Street, Toronto, Ontario, Canada M5S 1A7}
\author{Raymond Goerke}
\email{rgoerke@physics.utoronto.ca}
\affiliation{Department of Physics, University of Toronto, 60 St.\ George Street, Toronto, Ontario, Canada M5S 1A7}

\begin{abstract}
We calculate the anomalous dimensions of the next-to-leading order dijet operators in soft-collinear effective theory (SCET). We use a formulation of SCET where the Lagrangian is multiple copies of QCD and the interactions between sectors occur through light-like Wilson lines in external currents.  We introduce a small gluon mass to regulate the infrared divergences of the individual loop diagrams in order to properly extract the ultraviolet divergences.  We discuss this choice of infrared regulator and contrast it with the \dereg.  Our results can be used to increase the theoretical precision of the thrust distribution.
\end{abstract}

\maketitle


\section{Introduction\label{sec:intro}}

Perturbative calculations of jet observables involve multiple scales.  In the kinematic region where all the scales are much greater than $\lqcd$ but the ratio of these scales is small, often called the ``tail'' region, the rate is perturbative in both the strong coupling constant $\al_s$ and the ratio of the scales involved.  However, the rate includes large logarithms of the ratio of these scales at each order in perturbation theory.  These large logarithms limit the precision of theoretical predictions.  Effective field theory (EFT) techniques provide a framework to sum the terms enhanced by the logarithms using renormalization group equations (RGE).  This framework also contains a systematic procedure for including higher order effects in the small ratio of scales using subleading operators, allowing for logarithms suppressed by this small ratio to be summed in addition to those at leading order in the ratio.  These techniques can be used to improve the precision of the theoretical predictions.  In this paper we renormalize the next-to-leading order dijet operators in soft-collinear effective theory (SCET) with the purpose of using the RGE to sum the logarithms suppressed by the ratio of scales.  We will use the SCET operators introduced in the formulation of \cite{Freedman:2012,Freedman:2013vya}, in which the QCD dynamics of jets are described by multiple decoupled copies of QCD, and the EFT expansion only enters in the external currents.  Our results are useful for any observable requiring dijet operators; however, we will use the concrete example of the thrust rate to illustrate their usefulness.

Thrust \cite{Banfi:2001bz} is a useful jet shape observable for precision studies of high energy collisions, in particular for measuring $\al_s(M_Z)$ from LEP data\footnote{See \cite{Abbate:2010xh} and previous works by this collaboration.}.  Thrust is defined as
\beq
    \tau = 1-\frac{1}{Q}\max_{\vec{t}}\sum_{i\in X}\left|\vec{t}\cdot\vec{p}_i\right|,
\eeq
where $X$ is the final state, $Q$ is the total energy, and $\vec t$ is chosen to maximize the sum.  The integrated rate of the differential thrust distribution is defined by
\begin{align}\label{intrate}
    R(\tau)=\fr{1}{\si_0}\int_0^\tau\d\tau'\fr{\d\si}{\d\tau'},
\end{align}
where the Born rate is $\si_0$.  We will call this the thrust rate in the following.  A perturbative calculation of the thrust rate in the tail region where $(\La_\qcd/Q)\ll\tau\ll1$ involves three relevant scales: the hard scattering scale $Q$, the intermediate scale $\sqrt\tau Q$, and the soft scale $\tau Q$.  The rate can be written as an expansion $R(\tau)=R^{(0)}(\tau)+\tau R^{(1)}(\tau)+\ord(\tau^2)$ in this region, where the superscripts refer to the suppression in $\tau$, with $R^{(0)}$ and $R^{(1)}$ referring to $\ord(\tau^0)$ and \ordt\ rate respectively.   Each of the $R^{(i)}(\tau)$ terms in the thrust rate has an expansion in $\al_s$ of the form
\begin{align}\label{genrate}
    R^{(0)}(\tau)&=\sum_n\sum_{m\leq 2n}R^{(0)}_{nm}\al_s^n\log^m(\tau),\nn
    R^{(1)}(\tau)&=\sum_n\sum_{k\leq 2n-1}R^{(1)}_{nk}\al_s^n\log^k(\tau),
\end{align}
where the $R_{nm}^{(i)}$ are $\ord(1)$ constants and the large logarithms $\log\tau\gg1$ are due to the separation of scales.  The highest logarithmic power for the \ordt\ rate is suppressed by an additional power of $\al_s$ relative to the $\ord(\tau^0)$ rate.  When $\al_s\log\tau\sim\ord(1)$ the $\ord(\tau^0)$ rate becomes a divergent sum in increasing powers of $\log\tau$, spoiling the expansion in $\al_s(Q)\ll1$.  Although the \ordt\ rate has an overall suppression by $\tau$ compared to the $\ord(\tau^0)$ rate, the rate is similarly a sum in increasing powers of the logarithm.  Therefore, in order to restore a perturbative expansion in $\al_s$ for both the $\ord(\tau^0)$ and \ordt\ rates, the logarithms must be summed.

The $\ord(\tau^0)$ thrust rate has already been calculated to N$^3$LL accuracy and included the fixed order \ordt\ rate at $\ord(\al_s)$ \cite{Becher:2008cf}. In order to increase the theoretical precision in the tail region, the leading logarithms in the \ordt\ rate can become more important than further increasing the logarithmic accuracy in the $\ord(\tau^0)$ rate. Therefore, if the precision of the $\al_s(M_Z)$ measurement is to be improved, these former contributions to the thrust rate will need to be calculated.

The appropriate EFT for describing thrust is SCET \cite{Freedman:2012,\labelscet,Beneke:2002ni,Beneke:2002ph}.  SCET includes collinear and ultrasoft (usoft) fields that reproduce both the highly boosted and low energy degrees of freedom that are relevant in the tail region.  The expansion parameter of SCET is usually denoted by $\la$.  For thrust $\la\sim\sqrt{\tau}$, meaning the \ordt\ corrections require next-to-next-to-leading order in $\la$ (\nnlo) corrections to the effective theory\footnote{Unless otherwise stated, LO, \nlo, and \nnlo\ refer to the expansion in $\la$.}.  We use a formulation of SCET in which QCD fields are coupled to Wilson lines \cite{Freedman:2012}. Each of the sectors (usoft and collinear) interact amongst themselves via QCD, while the interactions between sectors are described by Wilson lines in appropriate representations.  This picture has been shown explicitly to \nnlo\ by doing a tree-level matching from QCD \cite{Freedman:2013vya}.  We contrast this formulation with the traditional approach to SCET, which has a Lagrangian expansion and mixes the various sectors \cite{\labelscet,Beneke:2002ni,Beneke:2002ph}.

SCET can sum the large logarithms in \eqref{genrate} by factorizing the rate and using the RGE to run from the hard scale to the soft scale.  The QCD operators are first matched onto the appropriate SCET dijet operators at the hard scale $Q$.  For the $\ord(\tau^0)$ rate we use the \lo\ dijet operators.  The \ordt\ rate requires the \nlo\ and \nnlo\ dijet operators, which are then run to the intermediate scale $\sqrt{\tau}Q$ using the RGE.  At the intermediate scale, the dijet operators are matched onto soft operators with the help of a factorization theorem.  The Wilson coefficients of the soft operators, often called the jet function, are run to the soft scale $\tau Q$.  The sequence of matching and running sums the large logarithms in the rate.

Recently, a factorization theorem has been shown for the \ordt\ rate \cite{Freedman:2013vya} that makes this possible.  The appropriate dijet operators and the tree-level matching coefficients were derived, as well as the appropriate soft operators.  By solving the RGE for the operators in \cite{Freedman:2013vya} the large logarithms in the \ordt\ rate can be summed.  In this paper we begin this process by calculating the anomalous dimensions of the \nlo\ dijet operators in SCET.  Summing all the logarithms in the \ordt\ rate of \eqref{genrate} also require the \nnlo\ dijet operators, which we leave for future work.

To compute the anomalous dimensions of the subleading effective operators we first compute their counterterms. We regulate using the $\msbar$ scheme and include a separate infrared (IR) regulator to ensure the $1/\ep$ poles are ultraviolet (UV) divergences.  The decoupling of the collinear and usoft sectors, manifest in the formulation of \cite{Freedman:2012}, means the IR cannot be regulated using a fermion off-shellness because the usoft sector will not be changed by this regulator.  We identify two possible IR regulators that will regulate the formulation of \cite{Freedman:2012}: the \dereg\ and a gluon mass.  The \dereg\ \cite{Chiu:2009yx} is similar to off-shellness but also modifies the Feynman rules of the usoft Wilson lines.  Unfortunately, the regulator introduces additional terms that make the calculation unnecessarily complicated.  We will demonstrate this in Section \ref{sec:deltareg}.  A gluon mass does not introduce any additional terms, meaning fewer calculations are needed.  However, this is done at the expense of introducing unregulated divergences in individual diagrams that only cancel if all the diagrams are added together before integrating. Either choice of regulator is equivalent since the counterterms do not depend on the IR regulator.  We chose to use a gluon mass.

The rest of the paper is organized as follows:  In Section \ref{sec:nlo} we briefly summarize the SCET formulation of \cite{Freedman:2012} and write the operators used in this calculation. We note that it was necessary to generalize the operators of \cite{Freedman:2012} in order to account for the mixing that occurs under renormalization. In Section \ref{sec:irreg} we discuss our choice of using a gluon mass as an IR regulator over the $\de$-regulator. We present the anomalous dimensions for the \nlo\ operators in Section \ref{sec:adnlo} and conclude in Section \ref{sec:concl}.


\section{SCET and NLO Operators \label{sec:nlo}}

In the kinematic region where thrust is dominated by collimated jets of light, energetic particles, SCET is the appropriate description.  It is convenient to introduce light-cone coordinates for describing the momentum of the highly boosted particles.  In light-cone coordinates the momentum is decomposed into two light-like components described by the vectors $n^\mu$ and $\nb^\mu$ as
\beq
    p^\mu=p\cdot n\frac{\nb^\mu}{2}+p\cdot \nb\frac{n^\mu}{2}+p_\perp^\mu.
\eeq
The vectors $n^\mu$ and $\nb^\mu$ satisfy $n^2=0=\nb^2$ and $n\cdot\nb=2$.  A boosted particle with $p\cdot\nb\sim Q$ will be described by \ncol\ fields in the effective theory.  Similarly, a boosted particle with $p\cdot n\sim Q$ will be described by an \nbcol\ field.  The perpendicular momentum of a collinear particle $p_{\perp}^\mu\sim\la Q$ is suppressed compared to the hard scale.  We must also include usoft fields that have no large components of momentum and whose momentum scales like $p^\mu\sim\la^2Q$.

We follow the approach of \cite{Freedman:2012} in deriving the \nlo\ SCET dijet operators.  Since particles in the same sector have no large momentum transfers, the interactions within each sector are governed by QCD.  Consequently, the Lagrangian has no expansion in $\la$ and can be written as
\beq
    \lag_\scet=\lag_\qcd^n+\lag_\qcd^\nb+\lag_\qcd^{us},
\eeq
where $\lag_\qcd^i$ is the QCD Lagrangian involving only $i^\mathrm{th}$-sector fields.\footnote{The approach of including decoupled copies of QCD for each sector has also been used to study factorization in QCD \cite{Feige:2013zla,Feige:2014wja}.} The interactions of particles in different sectors are described by external currents. Since these interactions involve large momentum transfers, the external currents can be organized into an expansion in $\la$. When computing the thrust rate in the limit $\tau\ll1$, the relevant external currents are dijet operators, which can be determined by matching the full QCD current
\beq\label{operatorexp}
    \bar\psi(x)\Ga\psi(x)= e^{-iQ(n+\nb)\cdot x/2}\left[\wc{0}\op{0}(x)+\fr1Q\sum_i\int\{\d\t\}\, C_2^{(1i)}(\{\t\})O_2^{(1i)}(x,\{\t\})+\ord(\la^2)\right]
\eeq
for a general Dirac structure $\Ga$.  The phase corresponding to the external momentum has been pulled out.  The superscripts in the dijet operators refer to the suppression in $\la$ and the $1/Q$ is included because the subleading operators are higher dimensional.  We have introduced a set of dimensionless shift variables $\{\t\}=\{Qt\}$ that were not included in \cite{Freedman:2012}; it will become apparent below that this shift corresponds to a displacement along a light-like direction describing the position of a derivative insertion. This generalization is needed in order to properly describe the mixing of operators under renormalization.

The leading order operator in \eqref{operatorexp} is \cite{Freedman:2012}
\beq\label{o20}%
    \op{0}(x)=\left[\bar\psi_n(\xnb)\wn{3}(\xnb,\xnbty)\right]\left[\yn{3}(\xsnty,0)\Ga_{(0)}\ynb{3}(0,\xsnbty)\right]\left[\wnb{3}(\xnty,\xn)\psi_\nb(\xn)\right],
\eeq%
and its matching coefficient is \cite{Manohar:2003vb,Bauer:2003}
\beq
    \wc{0}(\mu)=1+\fr{\al_s \cf}{4\pi}\left(-\log^2\fr{\mu^2}{-Q^2}-3\log\fr{\mu^2}{-Q^2}-8+\fr{\pi^2}{6} \right)
\eeq
where $\mu$ is the renormalization scale.  The Dirac structure is
\beq\label{Ga0}%
    \Ga_{(0)}=P_\nb\Gamma P_\nb
\eeq%
with projectors $P_n=(\nslash\bnslash)/4$ and $P_\nb=(\bnslash\nslash)/4$.  The subscripts on the fields denote the sector of the field. Each of the square brackets in \eqref{o20} are independently gauge invariant and corresponds to a separate sector.  The Wilson lines in the ${\bf R}$ representation
\begin{align}\label{wilsonline}%
    \wn{R}(x,y)&=P\exp\left(-ig\int_0^{n\cdot(y-x)/2}\d s \nb\cdot A_n^a(x+\nb s)T_{\bf R}^ae^{-s\epsilon}\right)\nn
    \yn{R}(x,y)&=P\exp\left(-ig\int_0^{\nb\cdot(y-x)/2}\d s n\cdot A_s^a(x+ns)T_{\bf R}^ae^{-s\epsilon}\right),
\end{align}%
represent a light-like colour source corresponding to the total colour of the other sectors (the symbol $P$ indicates path-ordering). The $\ep$ in the definition above gives the proper $i\ep$ pole prescription.  The $\wnb{R}$ and $\ynb{R}$ Wilson lines are defined similarly.  The positions in \eqref{o20}
\begin{align}\label{positions}%
    \xn   &=(0,x\cdot\nb,x_\perp)&   \qquad  \xnty   &=(0,\infty,x_\perp)\nn
    \xnb  &=(x\cdot n,0,x_\perp) &           \xnbty &=(\infty,0,x_\perp)\\
    \xsnty&=(\infty,0,0)         &           \xsnbty &=(0,\infty,0),\non
\end{align}%
come from multipole expanding the total momentum conservation constraint in $\la$ and is needed to ensure consistent power-counting at each order in $\la$.  

The \nlo\ operators are found by including $\ord(\la)$ corrections in the interactions between the sectors \cite{Freedman:2012}.  The operators that describe the modification to the \ncol\ sector are
\begin{align}\label{o1i}%
    \op{1a_n}(x,t)=&\left[\bar\psi_n(\xnb)\wn{3}(\xnb,\xnb+\nb t)iD^\mu_\perp(\xnb+\nb t)\wn{3}(\xnb+\nb t,\xnbty)\right]\nn
        \times&\left[\yn{3}(\xsnty,0)\Ga_{(1a_n)}^\mu\ynb{3}(0,\xsnbty)\right]\left[\wnb{3}(\xnty,\xn)\psi_\nb(\xn)\right]\nn
    \op{1b_n}(x,t)=&\left[\bar\psi(\xnb)\wn{3}(\xnb,\xnb+t\nb)iD_\perp^\mu(\xnb+t\nb)\wn{3}(\xnb+t\nb,\xnbty)\right]\nn
        \times&\left[\yn{3}(\xsnty,0)\Ga_{(1b_n)}^\mu\ynb{3}(0,\xsnbty)\right]\left[\wnb{3}(\xnty,\xn)\psi_\nb(\xn)\right]\nn
    \op{1B_n}(x)=&\left[\bar\psi(\xnb)\wn{3}(\xnb,\xnbty)i\overleftarrow{\partial}_\perp^\mu\right]\nn
        \times&\left[\yn{3}(\xsnty,0)\Ga_{(1b_n)}^\mu\ynb{3}(0,\xsnbty)\right]\left[\wnb{3}(\xnty,\xn)\psi_\nb(\xn)\right]\nn
    \op{1c_n}(x,t_1,t_2)=&\left[\bar\psi_n(\xnb)\wn{3}(\xnb, \xnb+t_2\nb)i\overleftarrow{D}_\perp^\mu(\xnb+t_2\nb)\wn{3}(\xnb+t_2\nb,\xnbty)\right]\nn
        \times&\left[\yn{3}(\xsnty, t_1n)i\overleftarrow{D}_\perp^\mu(t_1n)\yn{3}(t_1n,0)\Ga_{(1c_n)}\ynb{3}(0,\xsnbty)\right]\nn
        \times&\left[\wnb{3}(\xnty,\xn)\psi_\nb(\xn)\right]\nn
    \op{1d_n}(x,t)=&\left[ig\nb_\mu G_{n\perp}^{a\mu\nu}(\xnb)\wn{8}^{ab}(\xnb,\xnty)\right]\\
        \times&\left[\yn{8}^{bc}(\xsnty,tn)\bar\psi_s(tn)T^c\yn{3}(tn,0)\Ga_{(1d_n)}^\nu\ynb{3}(0,\xsnbty)\right]\nn
        \times&\left[\wnb{3}(\xnty,\xn)\psi_\nb(\xn)\right]\nn
    \op{1e_n}(x,t)=&\left[ig\nb_\mu G_{n\perp}^{a\mu\nu}(\xnb)\wn{8}^{ab}(\xnb,\xnty)\right]\left[\yn{3}^{d\hat d}(\xsnty,0)\ynb{8}^{\hat d c}(0,\xsnbty)\right]\nn
        \times&\left[\wnb{8}^{cb}(\xnty,\xn+tn)\bar\psi_\nb(\xn+tn)T^b\Ga_{(1e_n)}\wn{3}(\xn+tn,\xn)\psi_\nb(\xn)\right]\nn
    \op{1\de}(x)=&Q\left[\bar\psi_n(\xnb)x_\perp^\mu\wn{3}(\xnb,\xnbty)\right]\nn
        \times&\left[\yn{3}(\xsnty,0)\Ga_{(1\de)}(D_\perp^\mu+\overleftarrow{D}_\perp^\mu)(0)\ynb{3}(0,\xsnbty)\right]\left[\wnb{3}(\xnty,\xn)\psi_\nb(\xn)\right]\non,
\end{align}%
where the Dirac structures are
\begin{align}\label{g1i}%
    \Ga_{(1a_n)}^\mu=&P_\nb\Gamma\ga^\mu\fr{\nslash}2 &\qquad \Ga_{(1b_n)}^\mu =& \fr{\bnslash}2\ga^\mu \Ga P_\nb &\qquad \Ga_{(1c_n)}=&P_\nb\Gamma P_\nb \nn
    \Ga_{(1d_n)}^\mu=&\fr{\nslash}{2}\ga_\perp^\mu\Ga P_\nb  &\Ga_{(1e_n)}^\mu=&\fr{\nslash}{2}\ga^\mu_\perp\Ga P_\nb &\Ga_{(1\de)}=&P_\nb\Ga P_\nb.
\end{align}%
The covariant derivative is defined as $D^\mu(x)=\ptl^\mu-igT^aA^{a\mu}(x)$ and only couples the gluon to the corresponding sector on which it acts.  The field strength tensor is defined as $igG^{a\mu\nu}=f^{abc}[A^{b\mu},A^{c\nu}]$ where $f^{abc}$ are the $SU(3)$ structure constants.  The derivative in the $(1B_n)$ operator is strictly a partial derivative and not a covariant derivative because we are working in a covariant gauge where the gauge transformations at infinity vanish.  The $Q$ in front of the $(1\de)$ operator is required dimensionally.  The matching coefficients for the operators listed above are  \cite{Freedman:2012}
\begin{align}\label{c1i}%
    \wc{1a_n}(\t)   &=-\de(\t) +\oa              &\qquad \wc{1b_n}(\t)  &=\de(\t) +\oa \nn
    \wc{1B_n}       &=1+\oa                      &       \wc{1c_n}(\t_2,\t_1)&=2i\theta(\t_1)\de(\t_2) +\oa \\
    \wc{1d_n}(\t)   &=-2i\theta(\t) +\oa         &       \wc{1e_n}(\t) &=i\theta(\t) +\oa\nn
    \wc{1\de}(\t)   &=1+\oa, \non
\end{align}%
which are all dimensionless. The factors of $i$ ensure the convolution in \eqref{operatorexp} is real.

The \nlo\ operators explicitly decouple the sectors, just as in the \lo\ operator.  These operators differ from the \lo\ operators by a $D_\perp$ insertion at an arbitrary point along a Wilson line (for example the $(1a_n)$ operator) or by a change in the field content and Wilson line representation (for example the $(1e_n)$ operator).  The operators in \eqref{o1i} are generalizations of the \nlo\ operators in \cite{Freedman:2012,Freedman:2013vya}.  We find the form in \eqref{o1i} is necessary to properly renormalize the operators, since different values of the parameters can mix under renormalization.  We have also slightly changed the definition of the $(1b_n)$ operator and included the $(1B_n)$ operator, which makes the operator basis in \eqref{o1i} diagonal under renormalization.  

As was done in \cite{Freedman:2012}, we can compare the operators in \eqref{o1i} with the subleading operators in other formulations of SCET, such as in \cite{Hill:2004if}.  In \cite{Hill:2004if} the subleading heavy-to-light currents were renormalized.  While the dijet and heavy-to-light operators obviously differ in the usoft and \nbcol\ sectors, the modifications to the \ncol\ sector from the vector currents and subleading Lagrangian insertions in \cite{Hill:2004if} only differ from the corresponding operators in \eqref{o1i} by the appropriate Dirac structure basis.  This will serve as a way for us to compare the anomalous dimensions we calculate in Section \ref{sec:adnlo} with the results of \cite{Hill:2004if}.

We find it more convenient to work with the Fourier transformed operators $\opft{i}$ defined as
\begin{align}\label{fourier}%
    \opft{1i}(x,u)& = \int\dfthat e^{-iu\t}\op{1i}(x,\t) = Q\int\dft e^{-iQut}\op{1i}(x,t)\nn
    \opft{1i}(x,u_2,u_1)& = \int\dfthat[2]\dfthat[1] e^{-i(\t_2u_2+\t_1u_1)} \op{1i}(x,\t_2,\t_1).
\end{align}%
The matching in \eqref{operatorexp} is written in terms of these operators as
\beq\label{matchingtransform}
    \int\d\{\t\} \wc{1i}(\{\t\})\op{1i}(\{\t\}) = \int\d\{u\}\wcft{1i}(\{u\})\opft{1i}(\{u\})
\eeq
where
\begin{align}
    \wcft{1i}(u)& = \int\d\t \,e^{iu\t}\wc{1i}(\t) \nn
    \wcft{1i}(u_2,u_1) &= \int\d\t_2\d\t_1 \,e^{i(u_2\t_2+u_1\t_1)}\wc{1i}(\t_2,\t_1).
\end{align}
The $u$'s are momentum fractions at the vertex of the external current.  For collinear momentum $0\leq u\leq1$ due to momentum conservation, while for usoft momentum $0\leq u<\infty$ because usoft momentum is not conserved at the vertex.  The Fourier transformation of the NLO operators are
\begin{align}\label{o1iF}%
    \opft{1a_n}(x,u)&=\left[\bar\psi_n(\xnb)\de(u-in\cdot\hat D)iD^\mu_\perp(\xnb)\wn{3}(\xnb,\xnbty)\right]\nn
        &\times\left[\yn{3}(\xsnty,0)\Ga_{(1a_n)}^\mu\ynb{3}(0,\xsnbty)\right]\left[\wnb{3}(\xnty,\xn)\psi_\nb(\xn)\right]\nn
    \opft{1b_n}(x,u)&=\left[\bar\psi(\xnb)\de(u-in\cdot\hat D)iD_\perp^\mu(\xnb)\wn{3}(\xnb,\xnbty)\right]\nn
        &\times\left[\yn{3}(\xsnty,0)\Ga_{(1b_n)}^\mu\ynb{3}(0,\xsnbty)\right]\left[\wnb{3}(\xnty,\xn)\psi_\nb(\xn)\right]\nn
    \opft{1c_n}(x,u_2,u_1)&=\left[\bar\psi_n(\xnb)\de(u_2-in\cdot\hat D)i\overleftarrow{D}_\perp^\mu(\xnb)\wn{3}(\xnb,\xnbty)\right]\nn
        &\times\left[\yn{3}(\xsnty, 0)i\overleftarrow{D}_\perp^\mu(0)\de(u_1-in\cdot\overleftarrow{\hat D})\Ga_{(1c_n)}\ynb{3}(0,\xsnbty)\right]\nn
        &\times\left[\wnb{3}(\xnty,\xn)\psi_\nb(\xn)\right]\nn
    \opft{1d_n}(x,u)&=\left[ig\nb_\mu G_{n\perp}^{a\mu\nu}(\xnb)\wn{8}^{ab}(\xnb,\xnty)\right]\\
        &\times\left[\yn{8}^{bc}(\xsnty,0)\bar\psi_s(0)T^c\de(u-in\cdot\overleftarrow{\hat D})\Ga_{(1d_n)}^\nu\ynb{3}(0,\xsnbty)\right]\nn
        &\times\left[\wnb{3}(\xnty,\xn)\psi_\nb(\xn)\right]\nn
    \opft{1e_n}(x,u)&=\left[ig\nb_\mu G_{n\perp}^{a\mu\nu}(\xnb)\wn{8}^{ab}(\xnb,\xnty)\right]\left[\yn{3}^{d\hat d}(\xsnty,0)\ynb{8}^{\hat d c}(0,\xsnbty)\right]\nn
        &\times\left[\wnb{8}^{cb}(\xnty,\xn)\bar\psi_\nb(\xn)T^b\Ga_{(1e_n)}\de(u-in\cdot\overleftarrow{\hat D})\psi_\nb(\xn)\right]\non
\end{align}%
where $\hat D^\mu=D^\mu/Q$ is a dimensionless covariant derivative.  The tree-level matching coefficients up to $\ord(\al_s)$ corrections are
\begin{align}\label{c1iF}%
    \wcft{1a_n}(u)  &=-1                    &\wcft{1b_n}(u)  &=1 \nn
    \wcft{1c_n}(u_2,u_1)&=-\fr2{u_1}   &\wcft{1d_n}(u)&=\fr2{u} \qquad\qquad \wcft{1e_n}(u)  =-\fr1{u}
\end{align}%
The $(1B_n)$ and $(1\de)$ are independent of $\t$ so are not transformed.

\subsection{Constraining the NLO Operators}

We restrict ourselves to the electromagnetic current $\Ga=\ga^\la$ in this paper.  This current is both \cp\ invariant and conserved.  We will show how we can exploit these two properties to constrain the \nlo\ SCET operators.  We will also show how we can use the ambiguity in defining the $n^\mu$ and $\nb^\mu$ directions to make further constraints.  

First we use \cp\ invariance to expand the list of operators to include corrections to the \nbcol\ sector.  The action of \cp\ is equivalent to switching $n$ and $\nb$ and then taking the complex conjugate.  Therefore, the \nlo\ corrections to the \nbcol\ sector can be obtained for free from the operators in \eqref{o1i}.  The operators are
\begin{align}\label{1inb}%
    \op{1a_\nb}(x,t)=&\left[\bar\psi_n(\xnb)\wn{3}(\xnb,\xnbty)\right]\left[\yn{3}(\xsnty,0)\Ga_{(1a_\nb)}^\mu\ynb{3}(0,\xsnbty)\right]\nn
        &\times\left[\wnb{3}(\xnty, \xn+tn)i\overleftarrow{D}^\mu_\perp(\xn+nt)\wnb{3}(\xn+nt,\xn)\psi_\nb(\xn)\right]\nn
    \op{1b_\nb}(x,t)=&\left[\bar\psi_n(\xnb)\wn{3}(\xnb,\xnbty)\right]\left[\yn{3}(\xsnty,0)\Ga_{(1b_\nb)}^\mu\ynb{3}(0,\xsnbty)\right]\nn
        &\times\left[\wnb{3}(\xnty, \xn+tn)i\overleftarrow{D}^\mu_\perp(\xn+nt)\wnb{3}(\xn+nt,\xn)\psi_\nb(\xn)\right]\nn
    \op{1B_\nb}(x)=&\left[\bar\psi(\xnb)\wn{3}(\xnb,\xnbty)\right]\left[\yn{3}(\xsnty,0)\Ga_{(1b_\nb)}^\mu\ynb{3}(0,\xsnbty)\right]\nn
        &\times\left[i\ptl_{\perp\mu}\wnb{3}(\xnty,\xn)\psi_\nb(\xn)\right]\nn
    \op{1c_\nb}(x,t_2,t_1)=&\left[\bar\psi_n(\xnb)\wn{3}(\xnb,\xnbty)\right]\left[\yn{3}(\xsnty, 0)\Ga_{(1c_\nb)}\ynb{3}(0,t_1\nb)D_\perp^\mu(t_1\nb)\ynb{3}(t_1\nb,\xsnbty)\right]\nn
        &\times\left[\wnb{3}(\xnty,\xn+t_2n)iD_\perp^\mu(\xn+t_2n)\wnb{3}(\xn+t_2n,\xn)\psi_\nb(\xn)\right]\nn
    \op{1d_\nb}(x,t)=&\left[\bar\psi_n(\xnb)\wn{3}(\xnb,\xnbty)\right]\nn
        &\times\left[\yn{3}(\xsnty, 0)\Ga_{(1d_\nb)}^\nu\ynb{3}(0,t\nb)T^c\psi_s(t\nb)\ynb{8}^{bc}(tn,\xsnbty)\right]\\
        &\times\left[ign_\mu G_{\nb\perp}^{a\mu\nu}(\xn)\wnb{8}^{ab}(\xn,\xnbty)\right]\nn
    \op{1e_\nb}(x,t)=&\left[\bar\psi_n(\xnb)\wn{3}(\xnb,\xnb+t\nb)\Ga_{(1e_\nb)}T^b\psi_n(\xnb+t\nb)\wn{8}^{bc}(\xnb+t\nb,\xnbty)\right]\nn
        &\times\left[\yn{3}^{c\hat d}(\xsnty,0)\ynb{8}^{\hat d d}(0,\xsnbty)\right]\left[ign_\mu G_{\nb\perp}^{a\mu\nu}(\xn)\wnb{8}^{ad}(\xn,\xnbty)\right]\non
\end{align}%
with Dirac structures
\begin{align}\label{g1inb}%
    \Ga_{(1a_\nb)}^\mu&=\Ga_{(1b_n)}^\mu &\qquad \Ga_{(1b_\nb)}^\mu &= \Ga_{(1a_n)}^\mu &\qquad \Ga_{(1c_\nb)}&=\Ga_{(1c_n)} \nn
    \Ga_{(1d_\nb)}^\mu&= P_\nb\ga_\perp^\mu\Ga \fr{\nbslash}{2}  &\Ga_{(1e_\nb)}^\mu&=P_\nb\ga_\perp^\mu\Ga \fr{\nbslash}{2}.
\end{align}%
The $(1\de)$ is already \cp\ invariant since the $x_\perp^\mu$ can be moved into either collinear sector.  \cp\ invariance guarantees the matching coefficients of the $(1i_n)$ and $(1i_\nb)$ are equal
\beq
    \wc{1i_n}=\wc{1i_\nb}
\eeq
for $i=\{a,b,c,d,e,B\}$.  The Fourier transform of the operators in \eqref{1inb} are similar to those in \eqref{o1iF}, and we avoid writing them down for the sake of brevity.  In the following we will use \cp\ invariance to avoid talking about the $(1i_\nb)$ operators unless it is necessary.  

Next, we can exploit the conservation of the electromagnetic current $\ptl_\la \bar\psi(x)\ga^\la\psi(x)=0$.  As was discussed in \cite{Marcantonini:2008qn}, the EFT dijet operators must also be conserved at each order in $\la$ \footnote{We would like to thank Ilya Feige and Ian Moult for this observation}.  The only operators in \eqref{o1i} that are not conserved by themselves are the $(1a_n)$, $(1b_n)$, and $(1B_{(n,\nb)})$ operators.  All the other NLO operators are conserved up to $\ord(\la^2)$.  Therefore, conservation of the current requires
\beq\label{consconstraint}
    \wc{1a_n}=-\wc{1b_n}\qquad\wc{1B_n}=\wc{1B_\nb}
\eeq
to all orders in $\al_s$. 

Finally, we can exploit Reparameterization Invariance (RPI) \cite{Manohar:2002fd,Luke:1992cs} to constrain the matching coefficients.  RPI has been discussed extensively for heavy-to-light currents in the traditional SCET formulations \cite{Hill:2004if,Pirjol:2002km} but has not previously been discussed in the SCET formulation we use.  However, insight can be drawn from the traditional SCET formulations due to the equivalence of the two approaches.

The \ncol\ fields represent particles boosted in the $n^\mu$ direction, where $n^\mu$ is a vector we specify when matching from QCD onto SCET.  The \nbcol\ particles are described similarily. However, an \ncol\ particle does not travel exactly along the $n^{\mu}$ direction, and will generically have a momentum perpendicular to $n^{\mu}$ of order $\la$. In fact, we could have chosen a slightly different $n^{\mu}$, for example
\beq\label{rpin}%
    n'^\mu= n^\mu+\ep_\perp^\mu,
\eeq%
where $\ep_\perp\sim\ord(\la)$. In this case an \ncol\ particle also appears to be boosted along the $n'^{\mu}$ direction and has relative perpendicular momentum of order $\la$. Therefore, it should not matter to the physical result whether we include an \ncol\ sector or an $n'^{\mu}$-collinear sector. We can make use of this equivalence by applying the variation $n^{\mu}\to n^{\mu}+\ep_\perp^\mu$ to the operators in the \ncol\ sector and enforcing that they cancel order-by-order in $\la$. This provides constraints on the matching coefficients that must hold to all orders in $\al_s$.

Using the equation of motion for a Wilson line $n\cdot D\wn{R}=0$ and a fermion $\fmslash{D}\psi=0$, the variation of the \lo\ operator is
\begin{align}\label{rpinlo}
    &\op{0}(x)\xrightarrow{n^{\mu}\to n^{\mu}+\ep_{\perp}^{\mu}}\op{0}(x)\nn
    &+\left[\bar\psi_n(\xnb)\wn{3}(\xnb,\xnbty)\right]\left[\yn{3}(\xsnty,0)\de(\Ga_{(0)})\ynb{3}(0,\xsnbty)\right]\left[\wnb{3}(\xnty,\xn)\psi_\nb(\xn)\right]\nn
        &+\left[\bar\psi_n(\xnb)\fr{(\nb\cdot x)\ep_{\perp\mu}}{2}\wn{3}(\xnb,\xnbty)\right]\left[\yn{3}(\xsnty,0)\left(D^\mu+\overleftarrow{D}^\mu\right)\Ga_{(0)}\ynb{3}(0,\xsnbty)\right]\\
        &\quad\times\left[\wnb{3}(\xnty,\xn)\psi_\nb(\xn)\right]+\ord(\la^2),\non
\end{align}%
where
\beq\label{rpiga}%
    \de(\Ga_{(0)})=\fr{\bnslash}{2}\fr{\fmslash{\ep}_\perp}{2}\Ga P_\nb.
\eeq
Only the left projector is transformed because the Dirac structure is $\Ga_{(0)}=P_{\nb_1}\Ga P_{n_2}$ where $n_1^\mu$ and $n_2^\mu$ are the light-like directions of the two sectors.  However, matching enforces that $n\equiv n_1 = \nb_2$, so the transformed projector reduces to \eqref{rpiga}.

It is straightforward to show that the $(1\de)$ and $(1B_n)$ operators cancel the variations in \eqref{rpinlo} if their matching coefficients are constrained to be
\beq\label{rpiconstraint}%
    \wc{0}=\wc{1\de}=\wc{1B_n}
\eeq%
to all orders in $\al_s$.  We note this is similar to what was found in \cite{Hill:2004if} for heavy-to-light currents.  

We will check the constraints in \eqref{consconstraint} and \eqref{rpiconstraint} when we renormalize the NLO operators.  The anomalous dimensions, being the kernels of a linear integro-differential equation, are expected to be equal if two operators are constrained to be the same up to a multiplicative constant.  We will see this in Section \ref{sec:adnlo}.


\section{Infrared Regulator \label{sec:irreg}}

In order to extract counterterms from loop diagrams we must be able to differentiate between UV and IR divergences. Introducing a small mass scale to serve as an IR cut-off allows us to regulate the IR separately from dimensional regularization and ensures that all the $1/\ep$ poles in the loop integrals are UV divergences.  A common scheme is to introduce a small fermion off-shellness, as was done in \cite{Manohar:2003vb,Bauer:2000ew}.  However, in the SCET approach of \cite{Freedman:2012} where the sectors explicitly decouple, a fermion off-shellness leaves the Wilson lines unchanged and will not properly regulate the usoft sector of the \lo\ operator\footnote{In the traditional approach to SCET \cite{\labelscet} the \lo\ operator does not explicitly decouple until after a field redefinition, which does not affect the counterterms.}.  A regulator that produces similar results to a fermion off-shellness is the \dereg\ \cite{Chiu:2009yx}.  The \dereg\ modifies the Feynman rules of both the usoft and collinear sectors thereby regulating the IR of the SCET approach we use in this paper.  However, when there is more then one external leg in a single sector, the \dereg\ introduces extra terms that complicate the calculation.  Using a gluon mass to regulate the IR avoids these additional terms, although the individual diagrams will contain unregulated divergences, which cancel in the total sum of diagrams. We have chosen to use a gluon mass as our IR regulator, and in this section we will contrast some of the details of the two approaches.

In this and following sections we will use a condensed notation for representing the Feynman diagrams considered in our calculations. As an example to illustrate the notation, Figure \ref{fig:o20} shows the diagrams for \ncol\ quark and \nbcol\ anti-quark production using the LO dijet operator. This notation becomes especially useful when considering subleading operators with a gluon in the final state, as the number of diagrams grows considerably.

\bfig\begin{center}
    \subfloat[$I_n$]{$
        \bry{c}\bop[width=0.1\textwidth]{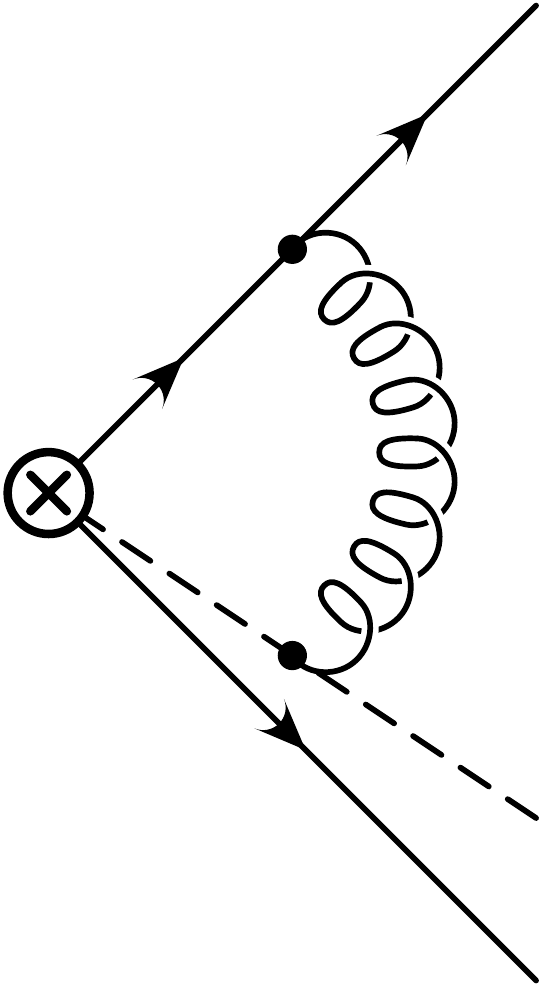}
            \put(35,97){$p_1$}
            \put(35,-1){$p_2$}
        \eop\ery
        =
        \bbm\bop[width=0.1\textwidth]{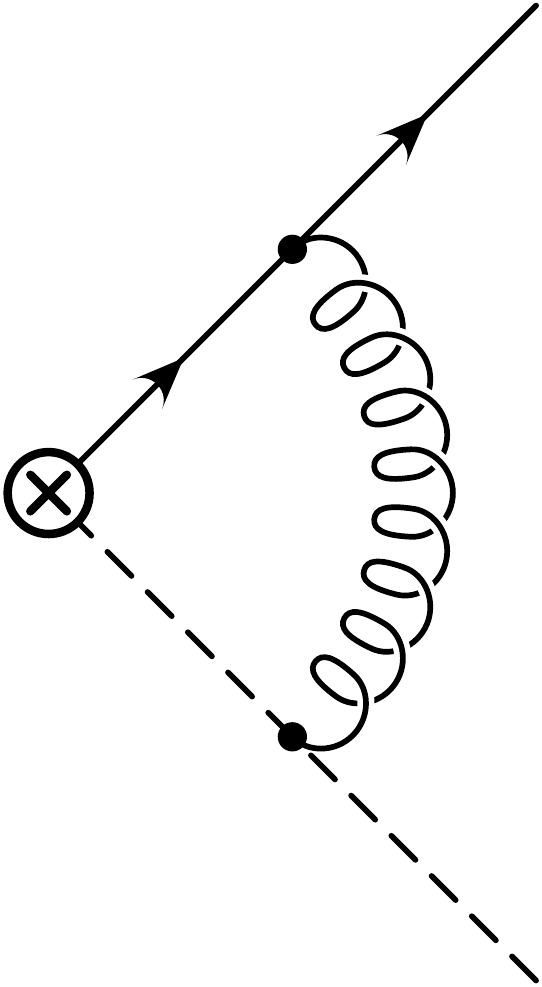}\put(35,97){$p_1$}\eop&\ebm
        \bbm\includegraphics[width=0.1\textwidth]{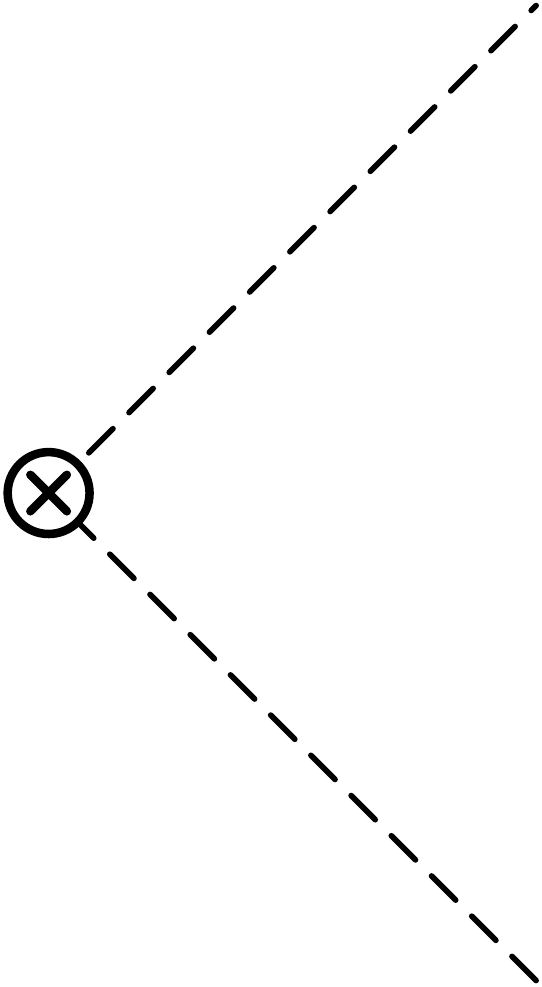}&\ebm
        \bbm\bop[width=0.1\textwidth]{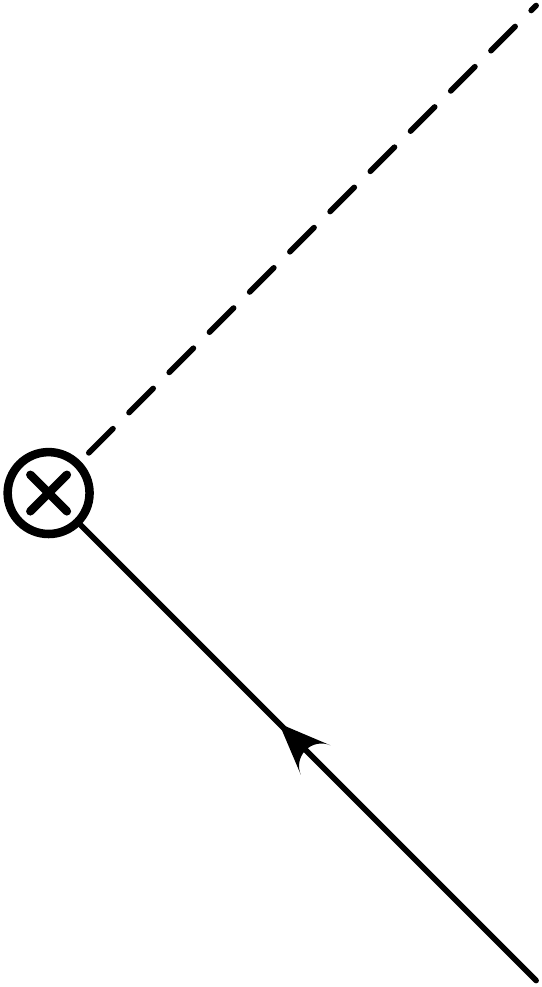}\put(35,-1){$p_2$}\eop&\ebm
        \equiv
        \bbm\includegraphics[width=0.1\textwidth]{0n.pdf}&\ebm
        $}
        \hspace{0.15\textwidth}
    \subfloat[$I_{us}$]{$
        \bry{c}\includegraphics[width=0.1\textwidth]{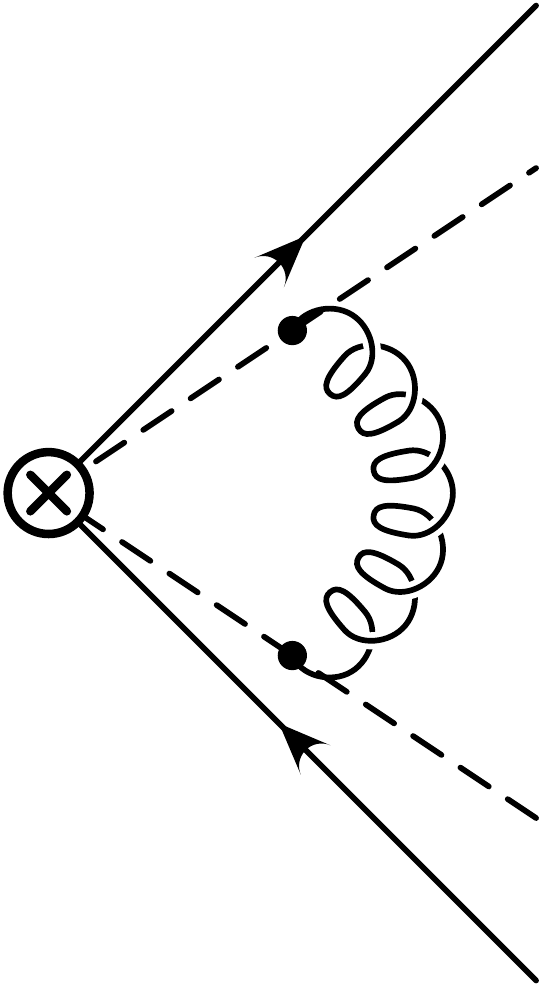}\ery=
        \bbm\includegraphics[width=0.1\textwidth]{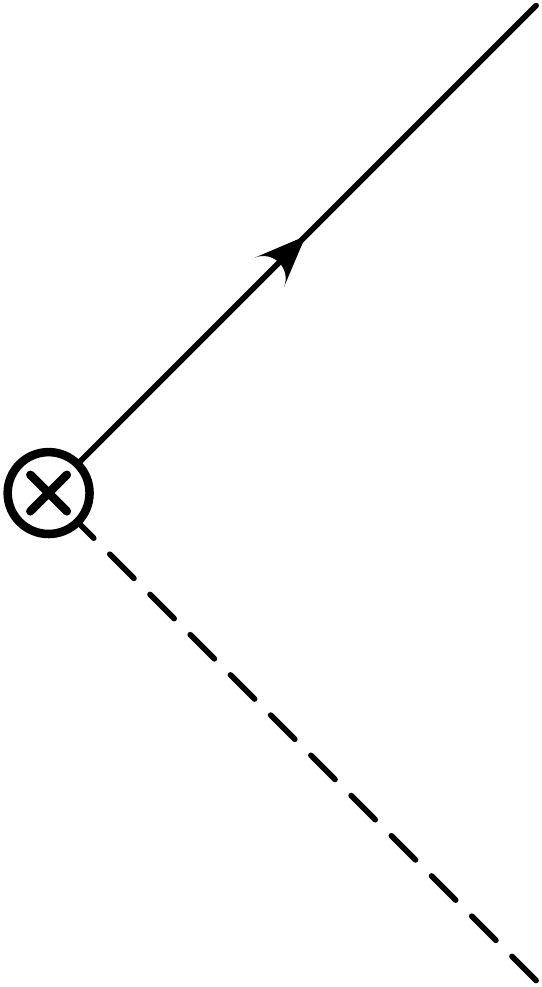}&\ebm
        \bbm\includegraphics[width=0.1\textwidth]{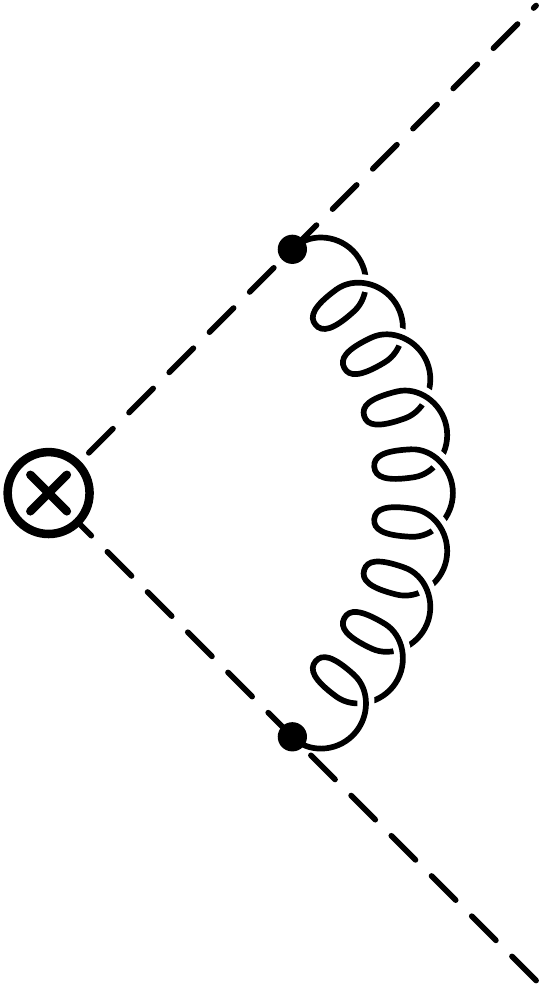}&\ebm
        \bbm\includegraphics[width=0.1\textwidth]{tree_nb.pdf}&\ebm
        \equiv
        \bbm\includegraphics[width=0.1\textwidth]{0s.pdf}&\ebm
        $}
        \hspace{0.15\textwidth}
    \subfloat[$I_\nb$]{$
        \bry{c}\includegraphics[width=0.1\textwidth]{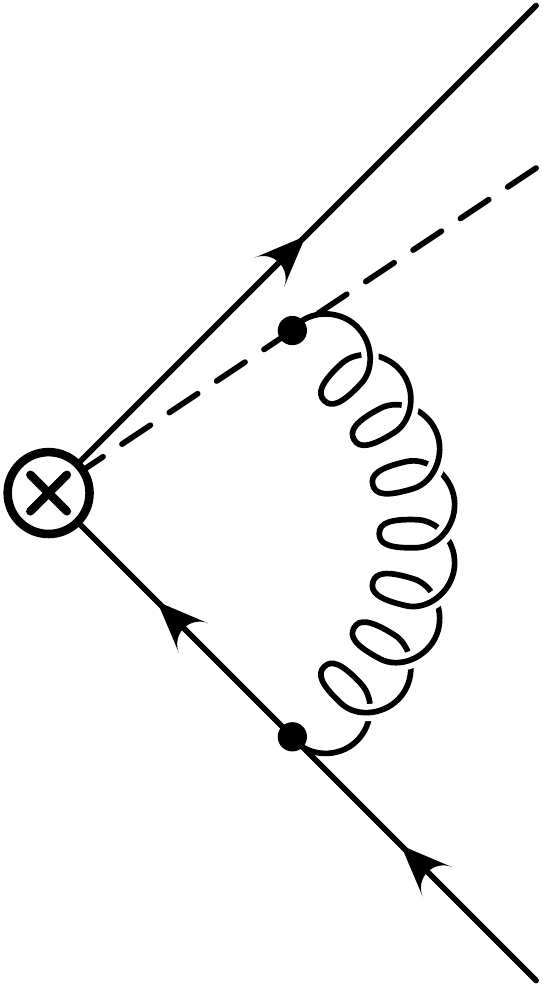}\ery=
        \bbm\includegraphics[width=0.1\textwidth]{tree_n.pdf}&\ebm
        \bbm\includegraphics[width=0.1\textwidth]{tree_s.pdf}&\ebm
        \bbm\includegraphics[width=0.1\textwidth]{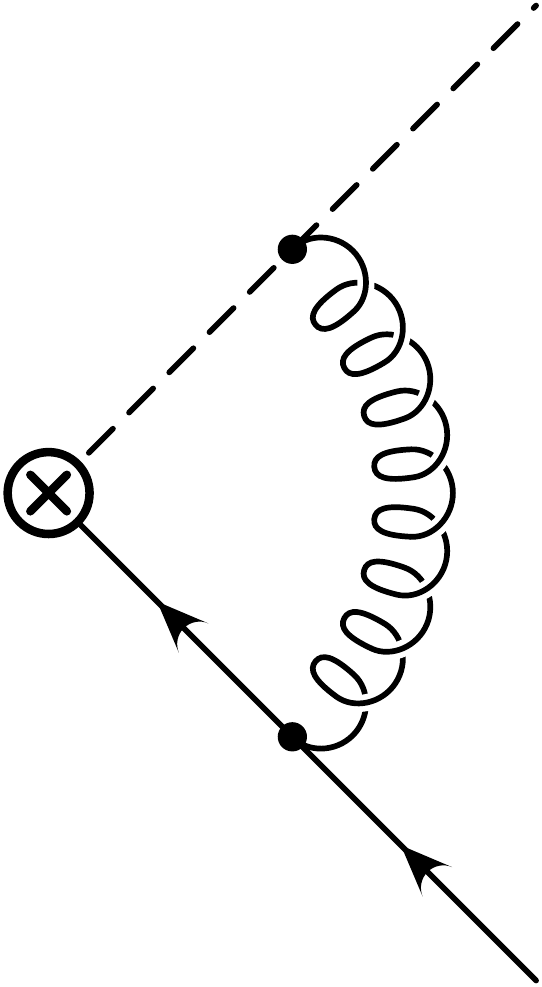}&\ebm
        \equiv
        \bbm\includegraphics[width=0.1\textwidth]{0nb.pdf}&\ebm
        $}
    \caption{\label{fig:o20} Relevant graphs for the renormalization of the $\op{0}$ operator. Solid lines and dashed lines are fermions and Wilson lines respectively.  We decompose the diagram on the left into the contribution from each sector in the middle three diagrams.  We can also compactify the notation by only showing the sector that has the one-loop contribution, as shown on the right.}
\end{center}\efig


\subsection{The Delta Regulator\label{sec:deltareg}}

The \dereg\ was introduced when considering SCET with massive gauge particles to help make the loop integrals of individual diagrams converge \cite{Chiu:2009yx}.  The construction is similar to using a fermion off-shellness and can be used to regulate the IR of the SCET formulation of \cite{Freedman:2012}.  This makes it an obvious choice for regulating the \nlo\ operators in \eqref{o1i}.  However, the \dereg\ requires extra terms when there is more than one external leg that do not appear when using a gluon mass to regulate the IR.

As an example to show where these extra terms arise, we renormalize the \lo\ dijet operator with an \nbcol\ anti-quark and an \ncol\ quark and gluon in the final state.  The diagrams are shown in Figure \ref{fig:nexternal}.  The \dereg\ regulates the IR by inserting a small mass term into the Lagrangian for each field.  The particles are brought off-shell by maintaining the massless equations of motion $p^2=0$.  The Feynman rules for the Wilson lines are also modified to incorporate this off-shellness.  The Feynman rules for the propagators and Wilson lines are \cite{Chiu:2009yx}
\beq\label{deltamod}
    \fr{1}{(p_i+k)^2-\De_i} \qquad \mathrm{and} \qquad  \fr{\nb_i^\al T_{{\bf R}j}^a}{k\cdot\nb_i-\de_{j,n}}
\eeq
respectively.  The momentum of the internal particle is $k$ and $\De_i$ is the mass term inserted into the Lagrangian.  The Feynman rule for the Wilson line is for a particle in the $j^\mathrm{th}$-sector with colour $T_{{\bf R}j}$ emitting a particle in the $i^\mathrm{th}$-sector\footnote{We note the colour structure was not in the original \dereg\ definition but is necessary when looking at $\ord(g^3)$ processes.}.  The shift in the Wilson line is $\de_{j,n}=(2\De_j)/((n_i\cdot n_j)(\nb_j\cdot p_i))$.  The regulator naively breaks the explicit decoupling into usoft and collinear fields.  However, once all the diagrams and their zero-bins have been accounted for, the result does factorize \cite{Chiu:2009yx}.

\bfig
    \subfloat[$I_n$]{$\bbm
        \includegraphics[width=0.1\textwidth]{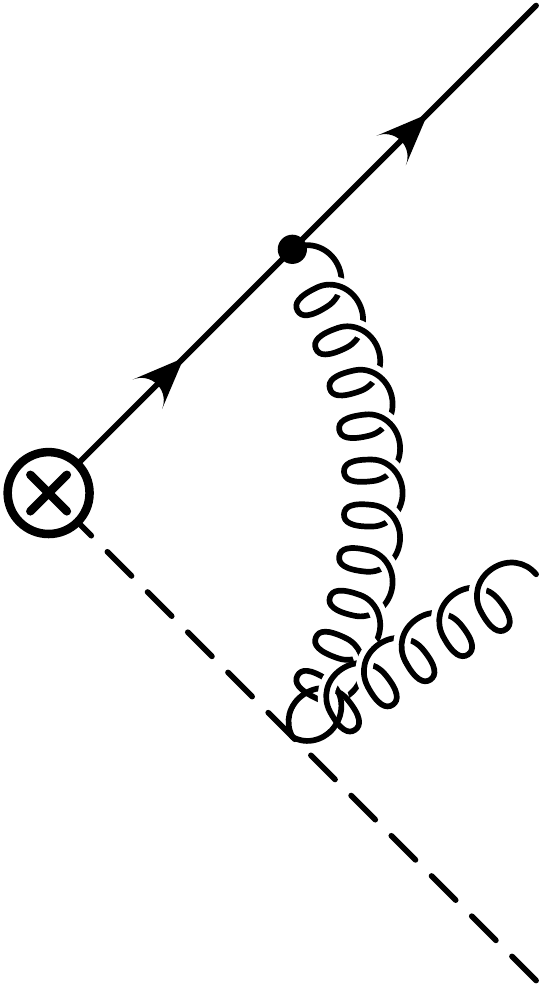} \hspace{0.01\textwidth}
        \includegraphics[width=0.1\textwidth]{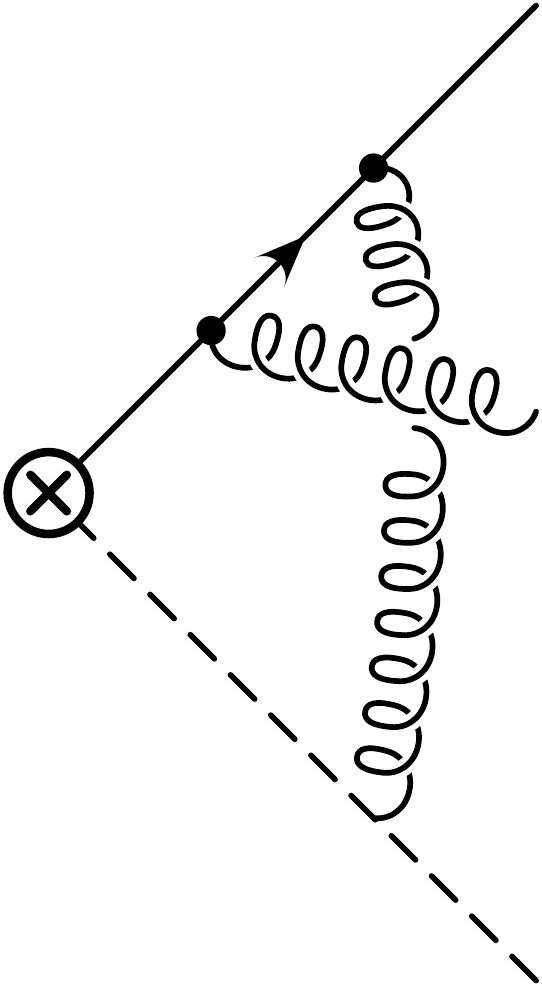} \hspace{0.01\textwidth}
        \includegraphics[width=0.1\textwidth]{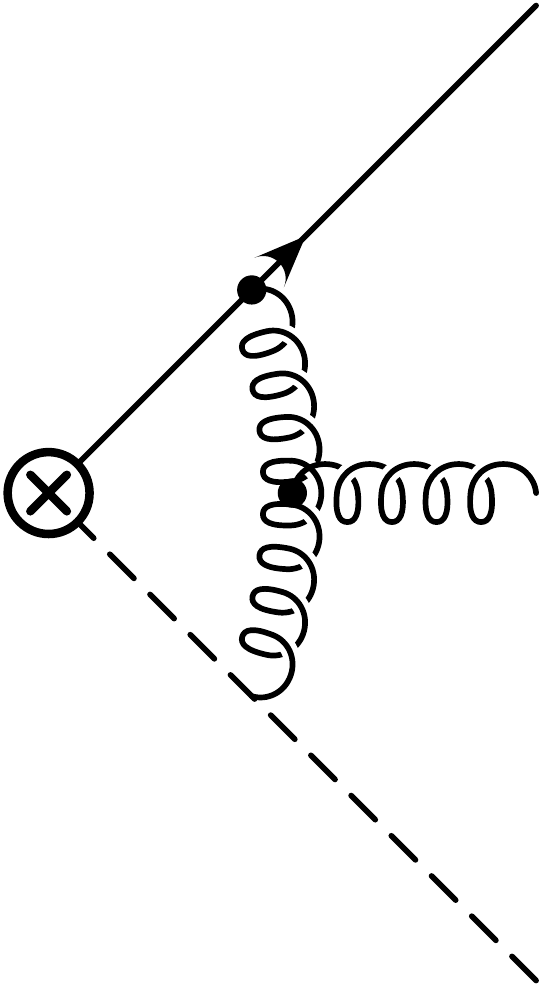} \hspace{0.01\textwidth}
        \includegraphics[width=0.1\textwidth]{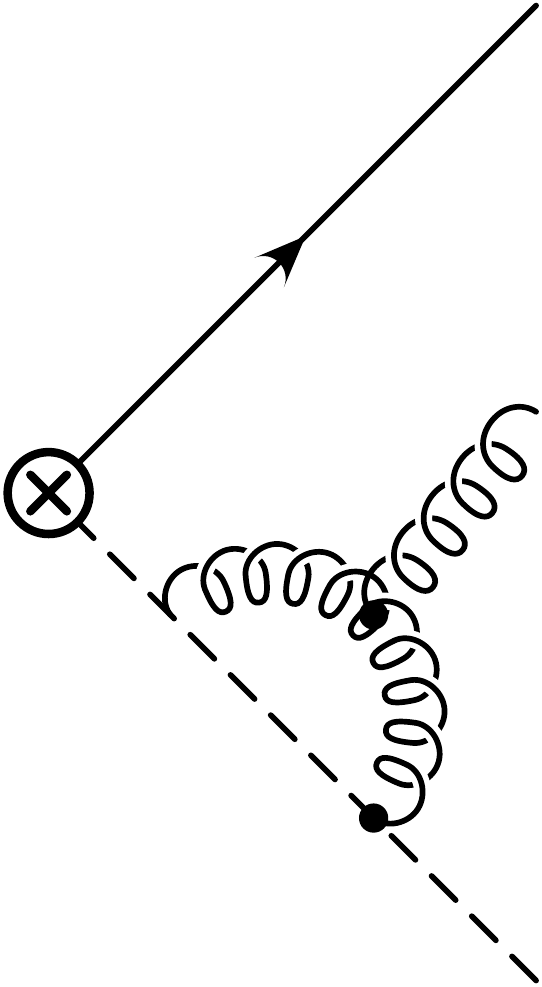} 
    \ebm$}
    \hspace{0.06\textwidth}
    \subfloat[$I_\nb$]{$\bbm\includegraphics[width=0.1\textwidth]{0nb}\ebm$} \hspace{0.06\textwidth}
    \subfloat[$I_{us}$\label{fig:iusoft}]{$\bbm\includegraphics[width=0.1\textwidth]{0s}\ebm$}
    \caption{One-loop diagrams for $\op{0}$ with an external \ncol\ gluon, using the compact notation of Figure \ref{fig:o20}.\label{fig:nexternal}}
\efig 

The modification to the Feynman rule of the Wilson line in \eqref{deltamod} leads to extra terms in the calculation of the diagrams.  For example, the usoft diagram in Figure \ref{fig:nexternal} leads to the integral
\beq\label{Is0delta}
    2ig^2\intfm k\frac{1}{(k^2-\De_g)(\nb\cdot k + \de_{\bar q,\nb})}\left( \fr{\cf+\ca}{n\cdot k-\de_{q,n}}-\fr{\ca}{n\cdot k-\de_{g,n}}  \right),
\eeq
where $\ka_\ep=(\mu^2e^{\ga_E})^\ep/(2\pi)^d$.  The extra $\ca$ terms account for the internal usoft gluon being emitted by or before the external \ncol\ gluon.  These extra terms are necessary to cancel all the mixed UV/IR divergences from the \ncol\ diagrams.  The \nbcol\ diagram will also require extra diagrams.  However, as expected, the final result reproduces the expected \lo\ anomalous dimension and is very similar to using a fermion off-shellness in a theory that does not decouple usoft and collinear fields.


\subsection{Gluon mass\label{sec:gluonmass}}

Another scheme that can be used to regulate the IR is to introduce a small gluon mass.  Unfortunately, massive bosons introduce an obstacle in SCET: the individual diagrams are often unregulated in dimensional regularization \cite{Chiu:2009yx}.  However, the sum of all the diagrams from a particular operator must still be well-regulated by a gluon mass \cite{Chiu:2009yx}.  As an example, we show how a gluon mass can be used to calculate the anomalous dimension of the \lo\ operator.  The necessary diagrams are shown in Figure \ref{fig:o20}. The \ncol\ diagram gives the integral
\begin{align}\label{In0}
    I_n &= 2ig^2 C_F \intfm k\frac{\nb\cdot(p_1-k)}{(k^2-M^2)(p_1-k)^2(\nb\cdot k)}\nn
        &= -2 g^2 C_F \ka_\ep \pi^{d/2}\Gamma(\ep)M^{-2\ep}\int_0^{p_1^-}\fr{\d\nbk}{\nbk}\left(1-\fr{\nbk}{p_1^-}\right)^{1-\ep}
\end{align}
where $M$ is the gluon mass.  The second line above is found by doing the $k^+$ integral by contours and then the $k_\perp$ integral.  The final integral diverges as $k^-\to0$ for all dimensions.  The usoft diagram gives the integral
\begin{align}\label{Is0}
    I_{us} &=2ig^2 C_F \intfm k \frac{1}{(k^2-M^2)(n\cdot k)(-\nb\cdot k)}\nn
        &=-2 g^2 C_F \ka_\ep \pi^{d/2}\Gamma(\ep)M^{-2\ep}\int_0^\infty\frac{\d\nbk}{\nbk}
\end{align}
after doing the same integrals as the \ncol\ diagram.  This integral diverges as $k^-\to0$ and $\infty$.  The \nbcol\ diagram gives the integral
\begin{align}\label{Inb0}
    I_\nb &=2ig^2 C_F \intfm k\frac{n\cdot(p_2-k)}{(k^2-M^2)(p_2-k)^2(\nb\cdot k)}\nn
          &= 2 g^2 C_F \ka_\ep \pi^{d/2}\Gamma(\ep)\left(\int_0^\infty \d\nbk\frac{p_2^+}{M^2+\nbk p_2^+}(-M^{-2\ep}+(-\nbk p_2^+)^{-\ep}) + \frac{M^{-2\ep}}{1-\ep}\right)
\end{align}
again doing the same integrals as the \ncol\ diagram.  The first term above diverges as $k^-\to\infty$.  As usual, we must also subtract a zero-bin $I_{n\zb}=I_{us}=I_{\nb\zb}$ for each of the collinear sectors \cite{Manohar:2007}.  Therefore, the sum of the diagrams is
\beq\label{scetint}%
    I_n+I_\nb-I_{us}.
\eeq
Each of the divergences in the above integrals cancel in the sum and we can find the anomalous dimension
\beq\label{ad0}%
    \ga_{2(0)}=\fr{\al_s\cf}{\pi}\left(\lq-\fr32\right).
\eeq%
This is the well-known result for the anomalous dimension of the \lo\ dijet operator \cite{Manohar:2003vb}.

Although the \dereg\ would avoid unregulated divergences in intermediate steps, it requires keeping track of additional terms. We chose to calculate the counterterms using a gluon mass and expect a \dereg\ to give the same results.


\section{Anomalous Dimensions \label{sec:adnlo}}

In order to run the \nlo\ Wilson coefficients in \eqref{c1i} from the high scale $Q$ to any other scale below $Q$, we must solve the RGE.  To do so we must renormalize the NLO operators and calculate their anomalous dimensions.

The renormalized operators $(R)$ and bare operators $(B)$ are related by
\beq\label{renorm}%
    \opft{1i}^{(B)}(\mu; x,\{u\})=\sum_j\int\{\d v\} \ct{1ij}(\mu ; \{u,v\})\opft{1i}^{(R)}( x,\{v\})
\eeq
where $\ct{1ij}$ is the counterterm matrix extracted from the UV divergences of the Green's functions of the operator.  In general, the continuous set of operators can mix within each label ${u}$ and with other operators $j$.  The independence of $\mu$ of the renormalized operators leads to an integro-differential equation for the bare operators
\beq
    \fr{\d}{\d\log\mu}\opft{1i}^{(B)}(\mu;x,\{u\})=-\sum_j\int\{\d v\}\ad{1ij}(\mu;\{u,v\})\opft{1j}^{(B)}(\mu;x,\{v\}).
\eeq
The anomalous dimension is calculated from the counterterms
\beq\label{anomdim}%
    \ad{1ij}(\mu ; \{u,v\})=-\sum_k\int \{\d w\} \ct{1ik}^{-1}(\mu ; \{u,w\})\fr{\d}{\d\log \mu}\ct{1kj}(\mu ; \{w,v\}).
\eeq%
The corresponding equation for the Wilson coefficients
\beq\label{rge}%
    \fr{\d}{\d\log\mu}\wcft{1i}(\mu ;\{u\})=\sum_j\int\{\d v\}\wcft{1j}(\mu ; \{v\})\ad{1ij}(\mu ; \{v,u\})
\eeq%
is the RGE that must be solved.

The operators in \eqref{o1iF} are written in a diagonal basis in $i,j$ up to $\oat$ corrections meaning 
\beq
    \ct{1ij}=
        \begin{cases}
            \ct{1i} & \textrm{if } i = j\\
            0   &   \textrm{if } i\neq j\,.
        \end{cases}
\eeq
The counterterms can be written perturbatively as
\beq\label{ct}%
   \ct{1i}(\mu ; \{u,v\})=\de(\{u-v\})+\alsb\ct{1i}^{(1)}(\mu ; \{u,v\})+\oat.
\eeq%
The anomalous dimension will also be diagonal in $i,j$ and the lowest order contribution will be
\beq\label{anomdimpert}%
    \ad{1i}(\mu ; \{u,v\})=2\left(\al_s\ep\fr{\ptl}{\ptl\al_s}-\fr{\ptl}{\ptl\log\mu^2}\right)\ct{1i}^{(1)}(\mu ; \{u,v\}).
\eeq%
The first term comes from the renormalization of the coupling constant $g^{(R)}=g^{(B)}\mu^{-2\ep}$. We will suppress the explicit dependence on $\mu$ in the anomalous dimension for the sake of more concise notation.

\afterpage{%
\begin{figure}[!ht]\centering
    \input{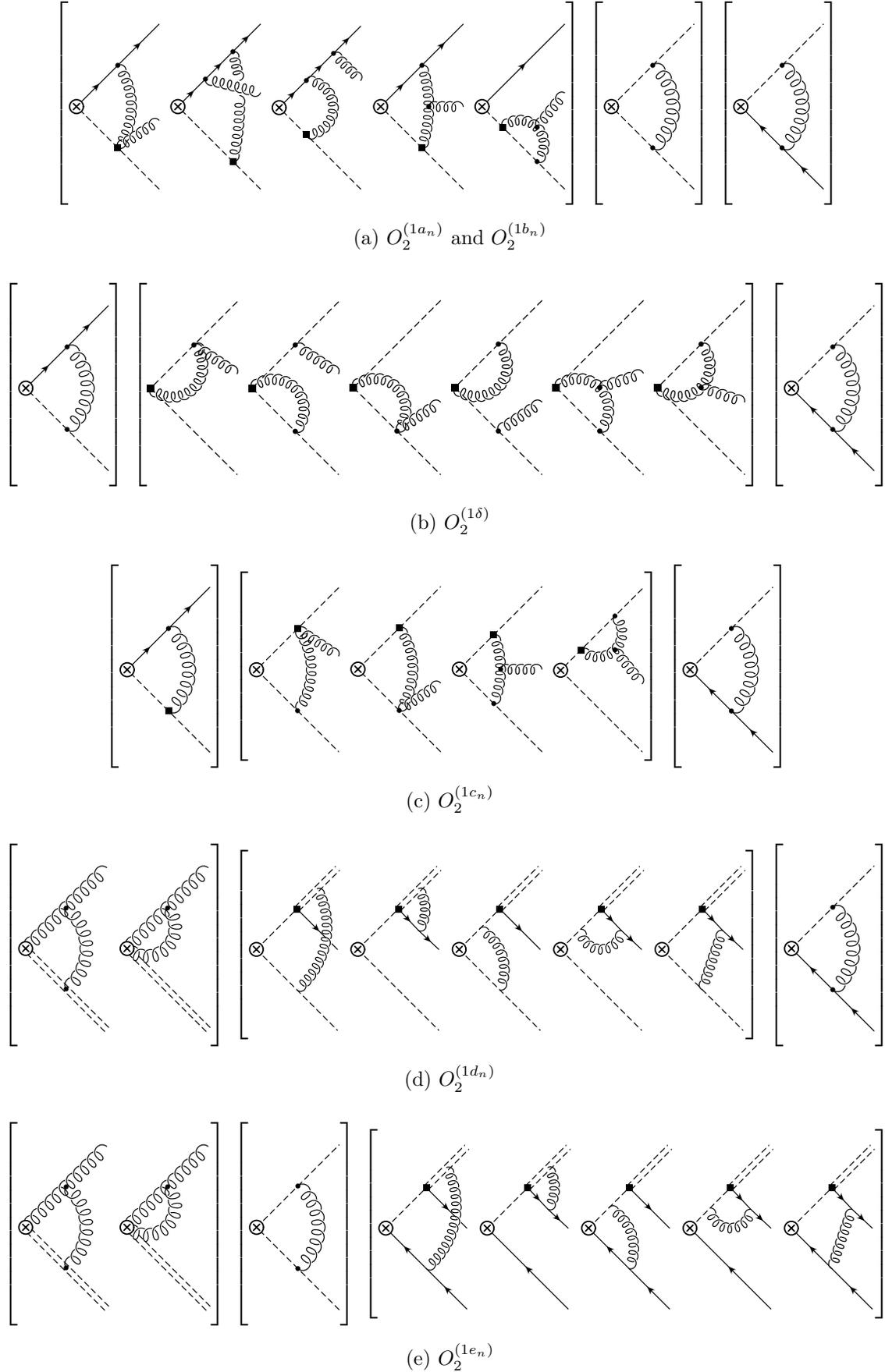}
    \caption{\label{fig:oi} Diagrams for the NLO operators.  Each bracket represents the one-loop graph from a separate sector.  Going from left to right, the diagrams are the \ncol, usoft, and \nbcol\ sectors.  The box vertex represents the derivative insertion. } 
\end{figure}
\clearpage}

The diagrams for the calculation of the anomalous dimensions of the \nlo\ operators are shown in Figure \ref{fig:oi}.  We must consider a gluon in the final state for most of the operators as these operators have a gluon in the final state at tree-level.  The $(1B_n)$ operator can be renormalized in a frame where the total perpendicular collinear momentum is non-zero and it has the same diagrams as the \lo\ operator in Figure \ref{fig:o20}.  We use the background field method \cite{Abbott:1980hw} to maintain gauge invariance under renormalization.  The background field method ensures $Z_g=Z_A^{-1/2}$, which properly renormalizes the derivative insertions and the Wilson lines.  Extracting the UV divergences from the diagrams lead to the following anomalous dimensions
\begin{align}\label{gamma1i}%
    \ga_{(1a_n)}(u,v)&=\fr{\al_s\de(u-v)\theta(\vb)}{\pi}\left(\cf\left(\lq-\fr32+\log\vb\right)+\fr\ca2 \right)\nn
        &+\fr{\al_s}{\pi}\left(\cf-\cah\right)\ub\left(\theta(1-u-v)\fr{uv}{\ub\vb}+\theta(\ub)\theta(\vb)\theta(u+v-1)\fr{uv+u+v-1}{uv} \right)\nn
        &-\fr{\al_s\ca}{2\pi}\ub\left( \theta(\ub)\theta(u-v)\fr{\vb-uv}{u\vb}+\theta(\vb)\theta(v-u)\fr{\ub-uv}{v\ub}\right.\nn
        &\left.+\fr1{\ub\vb}\PlusF{\fr{\ub\theta(\ub)\theta(u-v)}{u-v}+\fr{\vb\theta(\vb)\theta(v-u)}{v-u}}\right)\nn
    \ga_{(1b_n)}(u,v) &= \ga_{(1a_n)}(u,v)\nn
    \ga_{(1B_n)} &=\fr{\al_sC_F}{\pi}\left(-\fr32+\lq \right)=\ga_{(0)}\\
    \ga_{(1c_n)}(u_2;u_1,v_1) &=\fr{\al_s\de(u_1-v_1)\de(u_2)}{\pi}\left(\cf\left(-\fr32+\lq\right)+\fr\ca2\log v_1\right)\nn
        &-\fr{\al_s\ca\de(u_2)}{\pi}\left(\PlusF{\fr{\theta(v_1-u_1)\theta(u_1)}{v_1-u_1}+\fr{\theta(u_1-v_1)\theta(v_1)}{u_1-v_1}}\right.\nn
		&-\left.\fr{\theta(u_1-v_1)}{u_1}-\fr{\theta(v_1-u_1)}{v_1}\right)\nn
    \ga_{(1d_n)}(u,v) &=\fr{\al_s\de(u-v)}{\pi}\left(-\fr\cf2+\ca\left(\lq+\log(v)-\fr12\right)\right)\nn
        &-\fr{\al_s}{\pi}\left(\cf-\fr\ca2\right)\fr1v\PlusF{\fr{v\theta(u-v)\theta(v)}{u-v}+\fr{u\theta(v-u)\theta(u)}{v-u}}\nn
    \ga_{(1e_n)}(u,v) &=\fr{\al_s\de(u-v)\theta(\vb)}{\pi}\left(\fr{\cf}{2}+\ca\left(\lq+\log(v)-1\right)\right)\nn
        &-\fr{\al_s}{\pi}\left(\cf-\fr\ca2\right)\fr1{v\vb}\bigg(\theta(\vb)\theta(v-u)u\vb+\theta(\ub)\theta(u-v)v\ub\nn
		&\left.+\PlusF{\fr{\ub v\theta(\ub)\theta(u-v)}{u-v}+\fr{u\vb\theta(\vb)\theta(v-u)}{v-u}}\right),\non
\end{align}%
where $\ub=1-u$ and $\vb=1-v$.  We have used a generalized symmetric plus-distribution first introduced in \cite{Hill:2004if}, which was denoted by square brackets as in $\PlusF{\phantom{q}}$.  The formal definition of this distribution is
\begin{align} \label{genplus}
    &\left[ \theta(u-v)q(u,v) + \theta(v-u)q(v,u)\right]_+\nn
        &\equiv -\lim_{\beta\to 0} \frac{d}{du}\left[\theta(u-v-\beta)\int_u^{1+v}dw\,q(w,v) + \theta(v-u-\beta)\int^0_udw\,q(v,w)\right],
\end{align}
which is the same as the distribution defined in \cite{Hill:2004if} when $u, v\leq1$. The above definition is also valid when $u,v>1$, which was not required in \cite{Hill:2004if}.  Equation \eqref{gamma1i} is our main result.

We can compare the results for $\ad{1a_n}$ with \cite{Hill:2004if}.  The $(1a_n)$ operator in \eqref{o1i} is similar to the \nlo\ vector heavy-to-light current in \cite{Hill:2004if}.  As expected, the anomalous dimension for these two operators are the same for the non-diagonal terms.  They only disagree in the diagonal terms by the difference of the \lo\ dijet and heavy-to-light operator, which is expected.  Also, the anomalous dimensions of the $(1a_n)$ and $(1b_n)$ operators are the same, as expected from current conservation in \eqref{consconstraint}.  We can also check that $\ad{1i_\nb}=\ad{1i_n}$ as expected from \cp\ invariance. Finally, the $(1B_n)$, $(1\de)$, and $(0)$ operators all have the same anomalous dimension as expected from RPI.  

A final check is to compare the anomalous dimension of the $(1e_n)$ and $(1d_n)$ operators.  From \eqref{o1i} we see the $(1d_n)$ operator is the limit of the $(1e_n)$ operator when the quark becomes usoft.  Therefore, we expect in the limit where $u\sim\la^2\sim v$ in the $(1e_n)$ anomalous dimension to recover the $(1d_n)$ anomalous dimension.  This is indeed the case as seen in \eqref{gamma1i} \footnote{This limit must be taken carefully, since the $u\to\ord(\la^2)$ limit does not commute with the limit in the definition of the plus distribution.}.

The \nlo\ operators have a cusp in the usoft light-like Wilson lines at $x^\mu=0$.  Therefore, the anomalous dimension depends on at most a single logarithm and can be written in the form
\beq\label{cusp}
    \ad{1i}(\mu;u,v)=\de(u-v)\Gai{1i}^\mathrm{C}(\al_s)\lq+\gai{1i}^{\mathrm{NC}}(\al_s;u,v).
\eeq%
The coefficient of the logarithm is proportional to the universal cusp anomalous dimension $\Ga_\mathrm{cusp}(\al_s)$ \cite{Korchemsky:1987wg}, which means it is possible to perform NLL summation without going to higher loops.  This universal form of \eqref{cusp} is confirmed up to $\oat$ corrections in \eqref{gamma1i}.

Obviously, solving the RGE analytically is straightforward for the $(1B_n)$ and $(1\de)$ operators because the anomalous dimension is the same as the \lo\ dijet operator.  However, solving the RGE analytically for the other operators is more difficult.  The non-diagonal terms in the $(1a_n)$ and $(1b_n)$ RGE were solved in \cite{Hill:2004if} by exploiting that the non-diagonal terms in the anomalous dimensions can be written as $f(u,v)S(u,v)$ where $S(u,v)$ is a symmetric function.  For example, $f(u,v)=\ub$ for the $(1a_n)$ operator and $1/(v\vb)$ for the $(1e_n)$ operator.  The authors of \cite{Hill:2004if} were able to expand in an infinite set of Jacobi polynomials with the appropriate weight functions in order to diagonalize the anomalous dimensions and solve the RGE.  We expect that a similar solution will work for the $(1a_n)$, $(1b_n)$ and $(1e_n)$ operators.  However, the $(1c_n)$ and $(1d_n)$ operators are qualitatively different due to the limits on the labels, and a different strategy may be required. In any case, we believe it may be more practical to solve the RGE numerically, and we leave this for future work.


\section{Conclusion\label{sec:concl}}

In order to increase the accuracy of the $\al_s(M_Z)$ measurement the $\ord(\tau)$ corrections are becoming important.  Just like for the $\ord(\tau^0)$ rate, the \ordt\ rate includes large logarithms that must be summed.  We describe how this can be done using SCET and the factorization theorem in \cite{Freedman:2013vya}.  The required operators in the \ordt\ factorization theorem must be renormalized so they can be run from the hard scale to the usoft scale.  The running can be done in two stages.  First the \nlo\ and \nnlo\ dijet operators in SCET must be renormalized.  These operators are then run from the hard scale to the intermediate scale.  In the next step, the soft operators introduced in \cite{Freedman:2013vya} will be renormalized and run from intermediate scale to the usoft scale.  This sequence of running and matching will sum all the large logarithms in the \ordt\ rate.

In this paper, we have started the first step by renormalizing the \nlo\ dijet operators. Although we have used thrust as a concrete example of an application, our results is applicable to any observable requiring dijet operators.  Because we use the SCET formulation of \cite{Freedman:2012} we cannot use fermion off-shellness to regulate the IR. Instead we have used a gluon mass, which leads to individual diagrams being unregulated.  However, the sum of all the diagrams from a given operator is well-defined, as expected.  The UV divergences are extracted by looking at the $1/\ep$ poles allowing us to calculate the anomalous dimensions of the \nlo\ operators.  We have checked our results with similar operators for the heavy-to-light currents in \cite{Hill:2004if} and find good agreement.  

We leave renormalizing the N$^2$LO dijets operators and the soft operators to future work.  Although we have calculated the anomalous dimensions of the \nlo\ operators, and investigated the possibility of solving the RGE analytically, we believe that it may be more practical to solve it numerically, which we leave for future work.

\begin{acknowledgements}
We would like to thank Michael Luke, Ilya Feige, and Ian Moult for helpful discussions.  This work was supported by the Natural Sciences and Engineering Research Council of Canada.
\end{acknowledgements}

\bibliographystyle{JHEP}
\bibliography{renorm}

\providecommand{\href}[2]{#2}\begingroup\raggedright\begin{thebibliography}{10}

\bibitem{Freedman:2012}
S.~M. Freedman and M.~Luke, {\it {SCET, QCD and Wilson Lines}},  {\em
  Phys.Rev.} {\bf D85} (2012) 014003,
  [\href{http://xxx.lanl.gov/abs/1107.5823}{{\tt arXiv:1107.5823}}].

\bibitem{Freedman:2013vya}
S.~M. Freedman, {\it {Subleading Corrections To Thrust Using Effective Field
  Theory}},  \href{http://xxx.lanl.gov/abs/1303.1558}{{\tt arXiv:1303.1558}}.

\bibitem{Banfi:2001bz}
A.~Banfi, G.~Salam, and G.~Zanderighi, {\it {Semi-numerical resummation of
  event shapes}},  {\em JHEP} {\bf 0201} (2002) 018,
  [\href{http://xxx.lanl.gov/abs/hep-ph/0112156}{{\tt hep-ph/0112156}}].

\bibitem{Abbate:2010xh}
R.~Abbate, M.~Fickinger, A.~H. Hoang, V.~Mateu, and I.~W. Stewart, {\it {Thrust
  at $N^3LL$ with Power Corrections and a Precision Global Fit for
  $\alpha_s(m_Z)$}},  {\em Phys.Rev.} {\bf D83} (2011) 074021,
  [\href{http://xxx.lanl.gov/abs/1006.3080}{{\tt arXiv:1006.3080}}].

\bibitem{Becher:2008cf}
T.~Becher and M.~D. Schwartz, {\it {A precise determination of $\alpha_s$ from
  LEP thrust data using effective field theory}},  {\em JHEP} {\bf 0807} (2008)
  034, [\href{http://xxx.lanl.gov/abs/0803.0342}{{\tt arXiv:0803.0342}}].

\bibitem{Bauer:2000ew}
C.~W. Bauer, S.~Fleming, and M.~E. Luke, {\it Summing sudakov logarithms in
  $b\rightarrow{}x_s\gamma{}$ in effective field theory},  {\em Phys. Rev.}
  {\bf D63} (2000) 014006, [\href{http://xxx.lanl.gov/abs/hep-ph/0005275}{{\tt
  hep-ph/0005275}}].

\bibitem{Bauer:2000yr}
C.~W. Bauer, S.~Fleming, D.~Pirjol, and I.~W. Stewart, {\it An effective field
  theory for collinear and soft gluons: heavy to light decays},  {\em Phys.
  Rev.} {\bf D63} (2001) 114020,
  [\href{http://xxx.lanl.gov/abs/hep-ph/0011336}{{\tt hep-ph/0011336}}].

\bibitem{Bauer:2001ct}
C.~W. Bauer and I.~W. Stewart, {\it Invariant operators in collinear effective
  theory},  {\em Phys. Lett.} {\bf B516} (2001) 134--142,
  [\href{http://xxx.lanl.gov/abs/hep-ph/0107001}{{\tt hep-ph/0107001}}].

\bibitem{Bauer:2001yt}
C.~W. Bauer, D.~Pirjol, and I.~W. Stewart, {\it Soft collinear factorization in
  effective field theory},  {\em Phys. Rev.} {\bf D65} (2002) 054022,
  [\href{http://xxx.lanl.gov/abs/hep-ph/0109045}{{\tt hep-ph/0109045}}].

\bibitem{Bauer:2002nz}
C.~W. Bauer, S.~Fleming, D.~Pirjol, I.~Z. Rothstein, and I.~W. Stewart, {\it
  Hard scattering factorization from effective field theory},  {\em Phys. Rev.}
  {\bf D66} (2002) 014017, [\href{http://xxx.lanl.gov/abs/hep-ph/0202088}{{\tt
  hep-ph/0202088}}].

\bibitem{Beneke:2002ni}
M.~Beneke and T.~Feldmann, {\it {Multipole expanded soft collinear effective
  theory with non-Abelian gauge symmetry}},  {\em Phys. Lett.} {\bf B553}
  (2003) 267--276, [\href{http://xxx.lanl.gov/abs/hep-ph/0211358}{{\tt
  hep-ph/0211358}}].

\bibitem{Beneke:2002ph}
M.~Beneke, A.~Chapovsky, M.~Diehl, and T.~Feldmann, {\it {Soft collinear
  effective theory and heavy to light currents beyond leading power}},  {\em
  Nucl. Phys.} {\bf B643} (2002) 431--476,
  [\href{http://xxx.lanl.gov/abs/hep-ph/0206152}{{\tt hep-ph/0206152}}].

\bibitem{Chiu:2009yx}
J.-y. Chiu, A.~Fuhrer, A.~H. Hoang, R.~Kelley, and A.~V. Manohar, {\it
  {Soft-Collinear Factorization and Zero-Bin Subtractions}},  {\em Phys.Rev.}
  {\bf D79} (2009) 053007, [\href{http://xxx.lanl.gov/abs/0901.1332}{{\tt
  arXiv:0901.1332}}].

\bibitem{Feige:2013zla}
I.~Feige and M.~D. Schwartz, {\it {An on-shell approach to factorization}},
  {\em Phys.Rev.} {\bf D88} (2013), no.~6 065021,
  [\href{http://xxx.lanl.gov/abs/1306.6341}{{\tt arXiv:1306.6341}}].

\bibitem{Feige:2014wja}
I.~Feige and M.~D. Schwartz, {\it {Hard-Soft-Collinear Factorization to All
  Orders}},  \href{http://xxx.lanl.gov/abs/1403.6472}{{\tt arXiv:1403.6472}}.

\bibitem{Manohar:2003vb}
A.~V. Manohar, {\it {Deep inelastic scattering as $x\to 1$ using soft collinear
  effective theory}},  {\em Phys.Rev.} {\bf D68} (2003) 114019,
  [\href{http://xxx.lanl.gov/abs/hep-ph/0309176}{{\tt hep-ph/0309176}}].

\bibitem{Bauer:2003}
C.~W. Bauer, C.~Lee, A.~V. Manohar, and M.~B. Wise, {\it Enhanced
  nonperturbative effects in z decays to hadrons},  {\em Phys. Rev. D} {\bf 70}
  (Aug, 2004) 034014.

\bibitem{Hill:2004if}
R.~Hill, T.~Becher, S.~J. Lee, and M.~Neubert, {\it {Sudakov resummation for
  subleading SCET currents and heavy-to-light form-factors}},  {\em JHEP} {\bf
  0407} (2004) 081, [\href{http://xxx.lanl.gov/abs/hep-ph/0404217}{{\tt
  hep-ph/0404217}}].

\bibitem{Marcantonini:2008qn}
C.~Marcantonini and I.~W. Stewart, {\it {Reparameterization Invariant Collinear
  Operators}},  {\em Phys.Rev.} {\bf D79} (2009) 065028,
  [\href{http://xxx.lanl.gov/abs/0809.1093}{{\tt arXiv:0809.1093}}].

\bibitem{Manohar:2002fd}
A.~V. Manohar, T.~Mehen, D.~Pirjol, and I.~W. Stewart, {\it {Reparameterization
  invariance for collinear operators}},  {\em Phys.Lett.} {\bf B539} (2002)
  59--66, [\href{http://xxx.lanl.gov/abs/hep-ph/0204229}{{\tt
  hep-ph/0204229}}].

\bibitem{Luke:1992cs}
M.~E. Luke and A.~V. Manohar, {\it {Reparametrization invariance constraints on
  heavy particle effective field theories}},  {\em Phys.Lett.} {\bf B286}
  (1992) 348--354, [\href{http://xxx.lanl.gov/abs/hep-ph/9205228}{{\tt
  hep-ph/9205228}}].

\bibitem{Pirjol:2002km}
D.~Pirjol and I.~W. Stewart, {\it {A Complete basis for power suppressed
  collinear ultrasoft operators}},  {\em Phys.Rev.} {\bf D67} (2003) 094005,
  [\href{http://xxx.lanl.gov/abs/hep-ph/0211251}{{\tt hep-ph/0211251}}].

\bibitem{Manohar:2007}
A.~V. Manohar and I.~W. Stewart, {\it {The zero-bin and mode factorization in
  Quantum Field Theory}},  {\em Phys. Rev.} {\bf D76} (2007) 074002,
  [\href{http://xxx.lanl.gov/abs/hep-ph/0605001}{{\tt hep-ph/0605001}}].

\bibitem{Abbott:1980hw}
L.~Abbott, {\it {The Background Field Method Beyond One Loop}},  {\em
  Nucl.Phys.} {\bf B185} (1981) 189.

\bibitem{Korchemsky:1987wg}
G.~Korchemsky and A.~Radyushkin, {\it {Renormalization of the Wilson Loops
  Beyond the Leading Order}},  {\em Nucl.Phys.} {\bf B283} (1987) 342--364.

\end{thebibliography}\endgroup
\end{document}